\nocite{apsrev42Control}
\documentclass[aps,prb,reprint,longbibliography]{revtex4-2}
\usepackage[T1]{fontenc}
\usepackage[latin9]{luainputenc}
\usepackage{xcolor}
\usepackage{mathrsfs}
\usepackage{amssymb}
\usepackage{graphicx}
\usepackage{natbib}

\makeatletter

\@ifundefined{showcaptionsetup}{}{%
 \PassOptionsToPackage{caption=false}{subfig}}
\usepackage{subfig}
\makeatother

\usepackage{babel}
\begin{document}
\title{Low energy excitations in A-site ordered SmBaMn$_{2}$O$_{6}$}
\author{Mirian Garcia Fernandez$^{1*}$, Abhishek Nag$^{1,2}$, Stefano Agrestini$^{1}$,
Sahil Tippireddy$^{1}$, Dirk Backes$^{1}$, Xiaoyang Chen$^{1}$,
Urs Staub$^{3}$, Taka-hisa Arima$^{4}$ and Kejin Zhou$^{1}$}
\address{$^{1}$Diamond Light Source, Harwell Science and Innovation Campus,
OX11 0DE, Didcot, United Kingdom}
\address{$^{2}$Department of Physics, Indian Institute of Technology Roorkee,
Uttarakhand 247667, India}
\address{$^{3}$Center for Photon Science, Forschungsstrasse 111, Paul Scherrer
Institute, 5232 Villigen-PSI, Switzerland}
\address{$^{4}$Department of Advanced Materials Science, University of Tokyo,
Kashiwa 277-8561, Japan}
\begin{abstract}
The electron in a solid can be considered a bound state of the three
independent, fundamental degrees of freedom creating quasi-particles:
spinons, carrying the electron spin; plasmons carrying the collective
charge mode and orbitons carrying its orbital degree of freedom. These
fundamental degrees of freedom could form ordering states in which
dynamics or collective motions could occur and manifest as low-energy
excitations. The exotic properties that appear in the materials exhibiting
these electronic orderings are associated with these low-energy excitations.
Although the orbital order (OO) and its coupling to the spin system
creates very interesting phenomena, the microscopic origin of OO has
been much less explored than other electronic properties as it is
very difficult to directly access experimentally. Owing to the recent
improvement in energy resolution and flux, soft x-ray resonant inelastic
scattering (RIXS) allows for a re-examination of orbital excitations
in manganites. Here, we present a study of low energy excitations
in half doped A-site ordered SmBaMn$_{2}$O$_{6}$ through a combination
of RIXS and soft x-ray resonant elastic scattering (REXS) measurements.
We confirm the existence of OO at $\mathbf{q}$ = (0.25, 0.25, 0)
and find various low energy excitations below 200 meV. While several
excitations can be assigned to be of magnetic and phononic origin,
a group of excitations between 80 and 200 meV show a temperature dependence
closely following that of the OO making them possible candidates for
orbitons.
\end{abstract}
\maketitle

\section{Introduction}

Quasiparticles characterize low energy excitations in strongly correlated
electron systems. This concept is used to describe the collective
behavior by treating it as a single particle which is important to
simplify the many body problem. While fermionic quasiparticles have
been studied by angular resolved photoemission (ARPES), bosonic quasiparticles
can be probed by resonant inelastic scattering (RIXS). The orbital,
spin and charge degrees of freedom are quantum variables of electrons
and have thus their own dynamics. In the same way as a magnon being
a collective spin excitation in a magnetically ordered state; in a
long-range orbitally ordered system a collective orbital excitation
is predicted and its quantized object is known as orbiton.

Half doped manganites represent the prototype of orbitally ordered
systems. In these compounds half of the Mn ions are in the Mn$^{3+}$
state that has a d$^{4}$(t$_{2g}^{3}$ e$_{g}^{1}$) configuration
with one of the two e$_{g}$ orbitals being occupied by an electron
providing an orbital degree of freedom. Hole doping changes the ground
state due to the double-exchange interaction and at half doping orbital
order (OO) is formed\cite{Brink:1999aa}. Due to the strong electron-electron
correlation, the theoretical description of the manganite systems
is not straightforward. Even though some techniques, like Raman scattering\citep{Saitoh:2001aa}
and pump probe experiments\citep{Polli:2007aa} have claimed to have
observed collective orbital excitations in manganites, these results
are controversial\citep{Gruninger:2002aa} as these experiments cannot
provide information on the momentum dependence of orbitons. 

\begin{figure}
\subfloat[]{\begin{centering}
\includegraphics[scale=0.22]{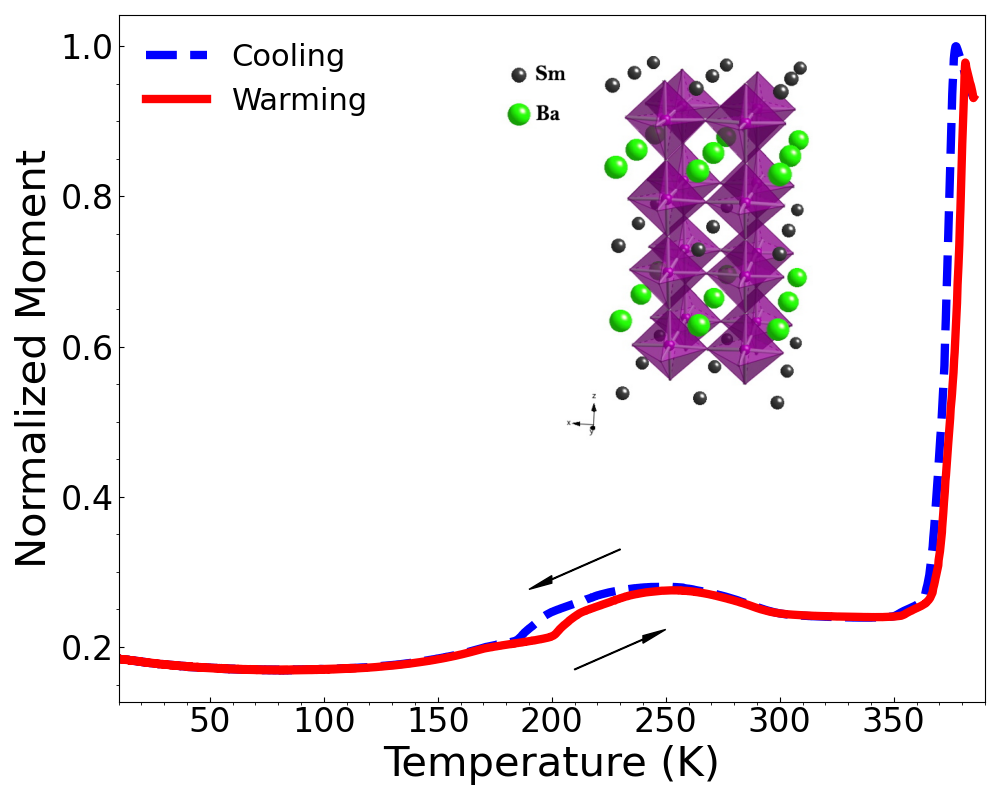}
\par\end{centering}
}

\subfloat[]{\centering{}\includegraphics[scale=0.25]{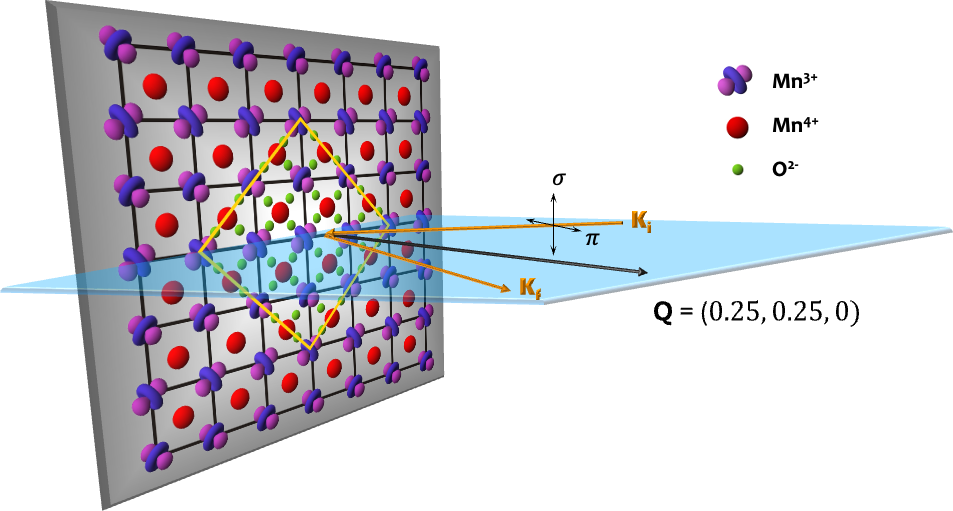}}\caption{\label{fig:Introduction}(a) Magnetization measured as function of
temperature. The insert shows the crystal structure of the A-site
ordered SBMO. (b) Experimental geometry, the blue plane represents
the horizontal scattering plane.}
\end{figure}

In the last years, resonant elastic x-ray scattering (REXS) has provided
much information about orbital physics by directly observing the orbital
order in different manganite compounds\citep{Murakami:1998aa,Wilkins:2003aa,Thomas:2004aa,Dhesi:2004aa,Staub:2005aa,M.Garcia-Fernandez2008,M.Garcia-Fernandez2009,Beale:2009aa,Staub:2009aa,M.Garcia-Fernandez2010}.
Several theoretical studies have predicted that orbital excitations
are distinguishable by characteristic variations in the RIXS amplitude
as a function of incident energy and momentum transfer\citep{PhysRevB.62.2338,ISHIHARA200415,PhysRevLett.101.106406}
from other excitations in the solid. These studies predict that both
single- and double-orbiton excitations are allowed, with intensities
that are of the same order. In one-dimensional cuprates, orbitons
are one of the three quasiparticles that electrons in solids are able
to split into during the process of spin-charge separation. In two-dimensional
cuprates, pure dispersive orbital excitations have been observed by
RIXS \citep{Schlappa:2012aa}. However, the detection and study of
the dispersion of orbitons in one of the prototypical orbital ordered
materials, i.e. half doped manganites has been elusive so far. To
our knowledge only \textit{K}-edge RIXS studies of doped manganites
can be found in the literature to date\citep{PhysRevB.64.014414,PhysRevB.67.045108,PhysRevB.70.224437,PhysRevLett.94.047203}.
As RIXS experiments at the \textit{K}-edge probe the 3d states indirectly\citep{PhysRevLett.94.047203},
studies at the Mn $L_{2,3}$-edge of doped manganites with high energy
resolution are required to bring light into this issue.

The half doped SmBaMn$_{2}$O$_{6}$ (SBMO) manganite crystallizes
in the slightly distorted perovskite structure ABO$_{3}$ depicted
in the insert of Figure \ref{fig:Introduction}(a). In this compound
the MnO$_{2}$ octahedra are sandwiched by two distinct layers built
of SmO and BaO\citep{doi:10.1143/JPSJ.73.2283} in contrast to the
disordered Sm$_{0.5}$Ba$_{0.5}$MnO$_{3}$ with a random distribution
of Sm and Ba ions. The A-site ordered \textit{R}BaMn$_{2}$O$_{6}$
compounds with \textit{R}=Sm, Eu, Gd, Tb, Dy, Ho and Y undergo a two-step
charge-orbital order (COO) transition at temperatures T$_{co1}$ =
380 K and T$_{co2}$ = 210 K hile this transition
is absent for the disordered compounds. Figure \ref{fig:Introduction}(a)
shows magnetization measurements of SBMO single crystal where these
transitions can be identified. Below T$_{co2}$ = 210 K, a disappearance
of the simple perovskite four-fold periodicity along the c-axis has
been suggested to be caused by the re-stacking of the COO planes.
The exact model, however, remains controversial\citep{doi:10.1143/JPSJ.71.2605,doi:10.1143/JPSJ.72.241,NAKAJIMA2004987,PhysRevB.66.140408,PhysRevB.70.064418}.
The issue of the unresolved models arises partly from the fact that
most experimental studies were performed on polycrystalline samples
which limits the direct observation of its microscopic physical properties.
A recent convergent beam electron diffraction experiment on a single
crystal of A-site ordered SBMO determined that the space group for
the T$_{co2-COO}$ and T$_{co1-COO}$ phases are \textit{P21am} and
\textit{Pnam }respectively\textit{ \citep{doi:10.1143/JPSJ.81.093602}.
}This space group change suggests that an even more complex structural
arrangement occurs below T$_{co2}$ as a consequence of the stacking
of the COO. Some theoretical studies even predict a ferroelectric
state in analogy to Pr(Sr$_{0.1}$Ca$_{0.9}$)$_{2}$Mn$_{2}$O$_{7}$
\citep{doi:10.1143/JPSJ.81.093602,Tokunaga:2006aa}. The existence
of this second charge order transition at 210 K, unique for the A-site
ordered manganites is responsible for a distinct temperature dependence
of the OO parameter that will be discussed later.

In this paper we present REXS and RIXS data which observe the orbital
order and its low energy excitations in a single crystal of SBMO.
We employ REXS in the vicinity of the Mn L$_{2,3}$ edges to compare
the OO signal to the previously published results on powder samples\citep{M.Garcia-Fernandez2008,M.Garcia-Fernandez2009}.
Using high-energy-resolution RIXS at these edges, we explore the existence
of orbitons and other low energy excitations.

Our results show that with the current available energy resolution
no orbiton could be detected in this compound at the OO ordering wave
vector. However, for deviating momentum transfers, several low energy
excitations could be observed that exhibit correlations with the compound's
electronic ordering phenomena.

We observed rich low-energy excitations up to \textasciitilde 200
meV at Mn $L_{3}$ and $L_{2}$ edges. Comparing to the results obtained
from Raman scattering and inelastic neutron scattering, the RIXS lower-energy
excitations comprise at least magnons and phonons. However, the integrated
spectral weight between 80 and 200 meV follows the distinct temperature
evolution of the OO peak in A-site ordered manganites, suggesting
the possible existence of orbitons in SBMO.

\section{Experimental details}

Single-crystalline SBMO was prepared in the following way: the mixed
powders were calcined at 1273 K in air, pressed into a rod, and sintered
at 1693 K under Ar atmosphere. At first, an A-site alloyed crystal
of Sm$_{1/2}$Ba$_{1/2}$MnO$_{3}$ was grown at a rate of 4--10
mm/h in air by a floating-zone method. The melt-quenched crystal was
treated in Ar atmosphere at 1693 K for 10 h to encourage the A-site
ordering, and then annealed in O$_{2}$ at 973 K for 2 h. The structural
characterization was performed by x-ray diffraction on a small part
of the powderized crystal. The Rietveld analysis with Rietan2000 indicated
the perfect ordering of the Sm and Ba atoms\citep{PhysRevB.70.064418}.
The magnetic and transport properties of the studied sample were characterized
using a SQUID magnetometer and are presented in Figure 1c. The crystal
orientation was determined by a lab-based Laue diffractometer prior
to the RIXS and REXS experiments.

The RIXS and REXS experiments were conducted at the I21-RIXS beamline
at Diamond Light Source, United Kingdom\citep{Zhou:rv5159}. The sample
was mounted with the {[}110{]} plane lying in the scattering plane
onto a copper sample holder fixed to a liquid He flow cryostat which
achieves temperatures between 7 K and 380 K. Experiments were performed
using linear horizontal or vertical polarized light leading to $\pi$
or $\sigma$ incident photon polarization in respect to the horizontal
scattering geometry. The REXS measurements were performed
using a photodiode as detector without the possibility to differentiate
the energy of the outgoing photons.

We tuned the incident photon energy to the resonance of the Mn L$_{2,3}$
absorption edges for the RIXS measurements. The total energy resolution
achieved was about 24 meV FWHM. RIXS and REXS signals were collected
without outgoing x-ray polarization analysis. For all RIXS spectra,
the elastic (zero-energy loss) peak positions were determined by the
elastic scattering spectrum from carbon tape placed near the sample
surface and then fine adjusted by its Gaussian fitted elastic peak
position. The Miller indices in this study are defined with reference
to the crystal structure $a_{p}$ x $a_{p}$ x 2$a_{p}$ with $a_{p}\thickapprox3.9$$\mathring{A}$
being the cubic perovskite unit cell. The experimental geometry of
the sample is depicted in Figure 1(d). The temperature of the measurements
was 15 K unless otherwise stated. The REXS measurements were performed
using a rotating photodiode inside the sample vessel. The RIXS measurements
were collected through continuous rotation of the spectrometer arm
due to the 3D nature of the investigated compound. The diffraction
peaks have been analyzed using Pseudo-Voigt functions while the inelastic
peak were described using Gaussian functions.

\section{Results and discussion}

\begin{figure}
\begin{centering}
\includegraphics[scale=0.22]{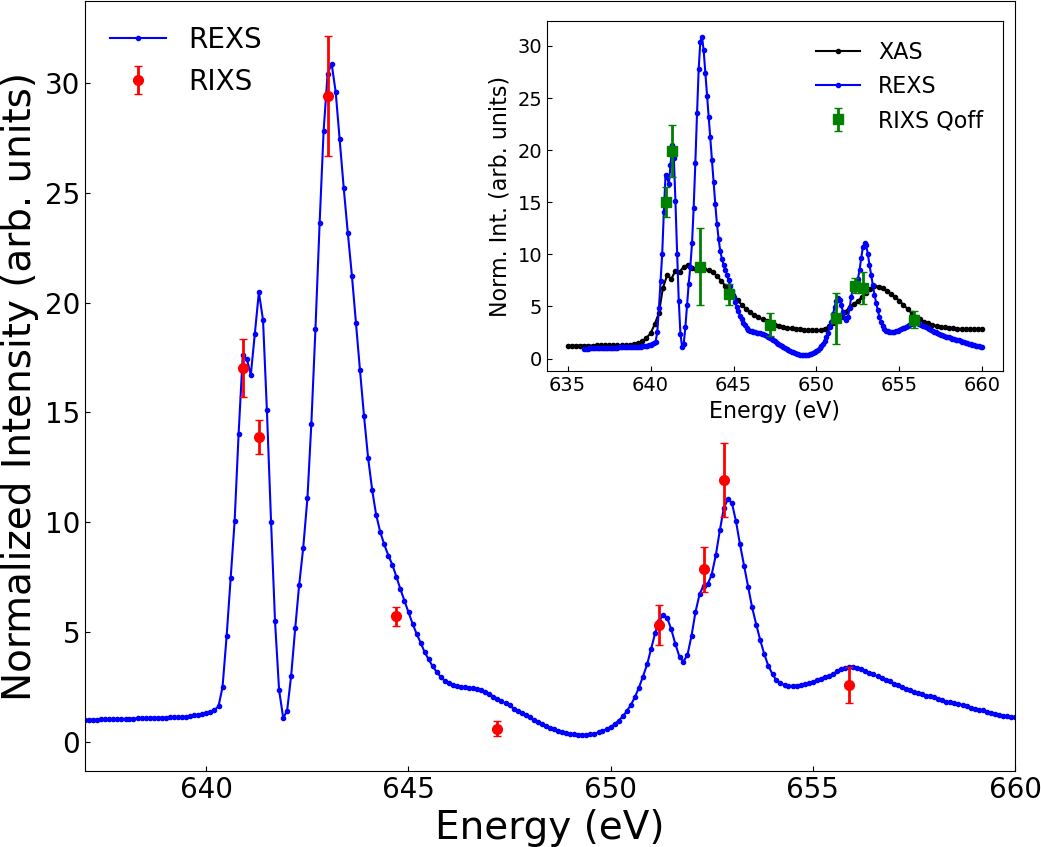}
\par\end{centering}
\caption{\label{fig:Energy-dependence}Energy dependence of the (0.25, 0.25,
0) OO reflection together with the integrated intensity of the RIXS
signal collected for different incident energies while maintaining
the momentum transfer $\mathbf{q}$ constant. The insert shows the
XAS taken in fluorescence mode together with the energy dependence
of the Q$_{off}$ reflection. The measurements were performed with
$\sigma$ incident polarization at T = 15K.}
\end{figure}

Figure \ref{fig:Energy-dependence} shows the REXS data of energy
dependence of the orbital (1/4 1/4 0) reflection in the vicinity of
the Mn $L_{2,3}$ edges. The insert of Figure \ref{fig:Energy-dependence}
shows the comparison to the experimental x-ray absorption (XAS). Clearly,
the resonant reflection intensity across the Mn $L_{2,3}$ edges is
distinct from the XAS signal and evidences the OO \citep{M.Garcia-Fernandez2008,M.Garcia-Fernandez2009,M.Garcia-Fernandez2010}.
More specifically, the present measurements on SBMO shows the same
differences with the layered La$_{0.5}$Sr$_{1.5}$MnO$_{4}$ spectra
observed previously in literature \citep{M.Garcia-Fernandez2008}.
These data confirm that the OO of the e$_{g}$ electrons is of x$^{2}$-z$^{2}$/y$^{2}$-z$^{2}$
type in A-site ordered SBMO, while a 3x$^{2}$-r$^{2}$/3y$^{2}$-r$^{2}$
type of orbital order is present in the layered manganite.

Following the REXS measurements, we performed the energy dependence
RIXS measurements with the $\theta$ and 2$\theta$ angles ensuring
a fixed momentum transfer at the (1/4 1/4 0) reflection when varying
the x-ray energy. The integrated intensities between
$\pm$ 80 meV of these RIXS spectra are shown as red dots in Figure
2 together with the REXS signal. Both RIXS and REXS signals are very
similar confirming the excellent alignment of the spectrometer to
$\mathbf{q_{OO}}$= (0.25, 0.25, 0). In order to
minimize the elastic signal, the RIXS measurements shown later were
collected both at $\mathbf{q_{OO}}$ and off from the OO Bragg condition
at $\mathbf{q}_{Off}=(0.2347,0.2347,-0.0130)$. In the insert of Figure
2 the comparison with the observed energy dependence measured at fixed
$\mathbf{q_{Off}}$ with the REXS and the XAS signal is shown.

\begin{figure}
\includegraphics[scale=0.2]{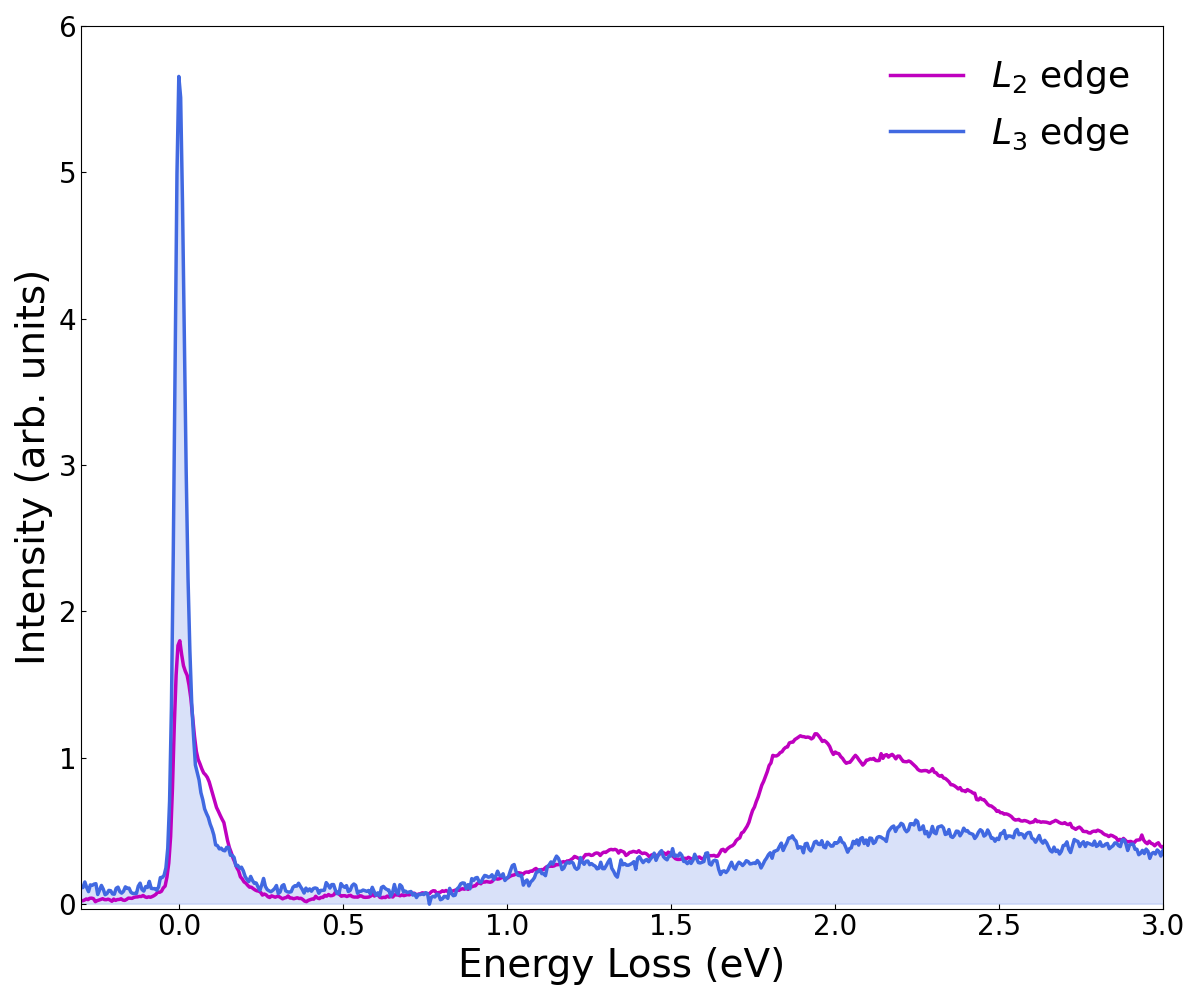}

\caption{\label{fig:Low-energy-transfer-RIXS}Low energy-transfer RIXS line
spectrum measured at the Mn $L_{2}$ and $L_{3}$ edge with $\sigma$
incident polarization and T=15K plotted in different axis. The $\protect\overrightarrow{q}$
vector for these measurements corresponds to (0.2347, 0.2347, -0.0130)}
\end{figure}

\textcolor{black}{Figure \ref{fig:Low-energy-transfer-RIXS} shows
the difference of spectral weight between the measurements done at
the $L_{3}$ and $L_{2}$ edges. Due to the self absorption, the $L_{2}$
RIXS spectra shows more clearly the d-d and low energy excitations,
whereas the $L_{3}$ spectra is mainly dominated by the elastic signal.
Keeping this information in mind we proceeded to study the low-energy
excitations }at the incident energy of 652.85 eV in order to explore the potential existence of the collective orbiton. 

\begin{figure*}
\subfloat[]{\begin{raggedright}
\includegraphics[scale=0.22]{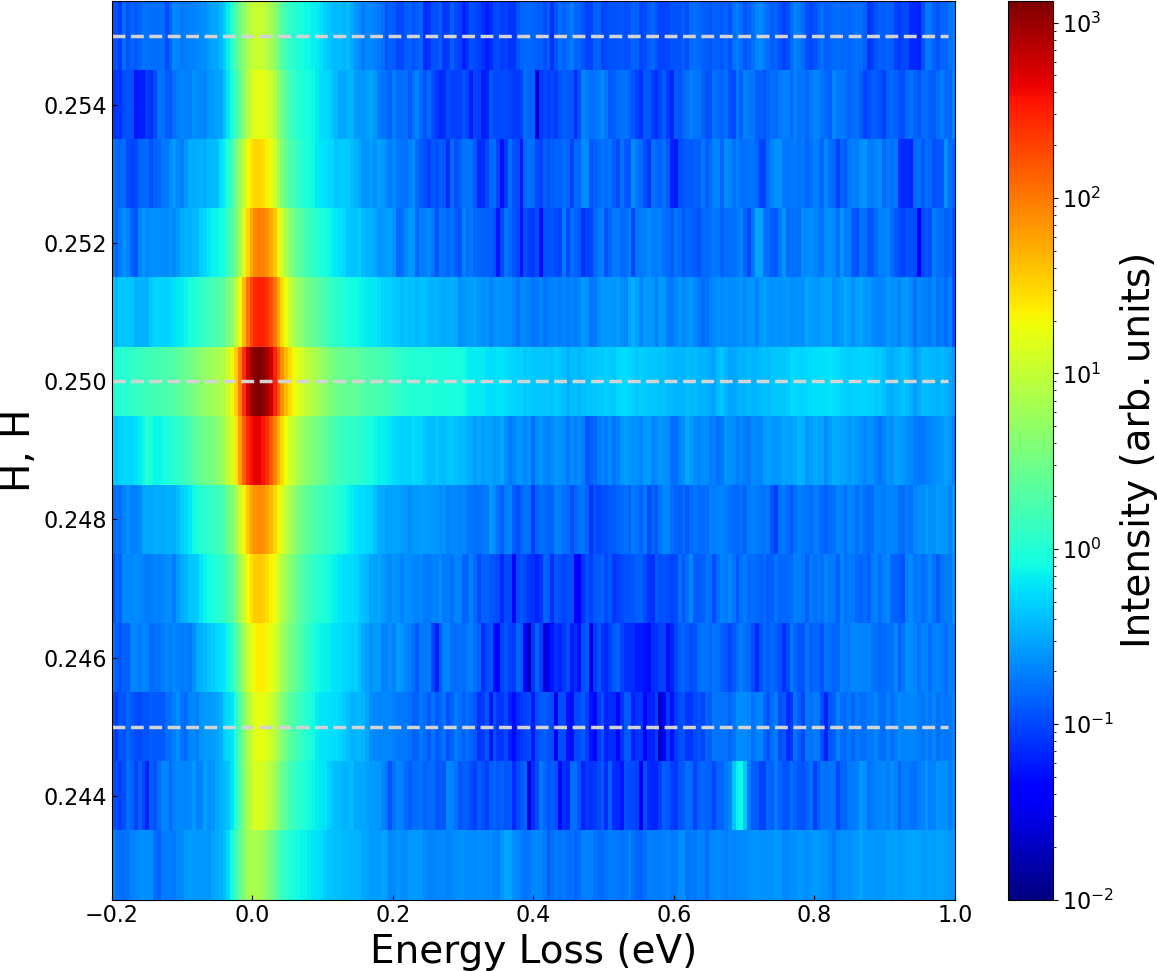}
\par\end{raggedright}
}\subfloat[]{\includegraphics[scale=0.22]{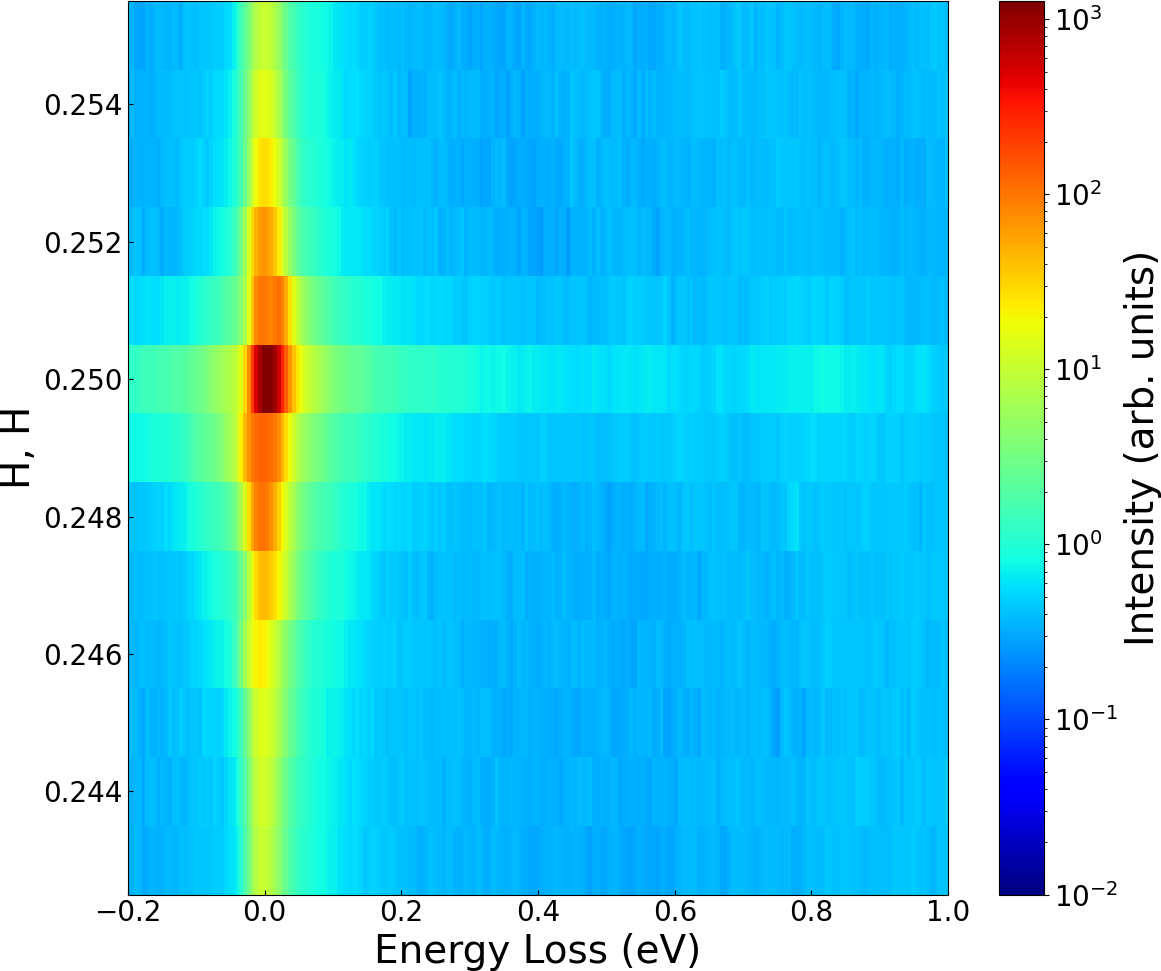}}

\caption{\label{fig:q-maps}Low energy excitations of SBMO measured in the
vicinity of the OO (0.25, 0.25, 0) reflection revealed by Mn $L_{2}$
RIXS and T = 15K. (a) Incident $\pi$ polarization. The
feature at HH 0.244 and energy loss 0.7 eV corresponds to a cosmic
ray in the detector.} (b) Incident $\sigma$ polarization.
\end{figure*}

We now discuss the inelastic excitation at the OO $\mathbf{q}$$_{OO}$
=(0.25, 0.25, 0). Given that the inelastic signal is stronger at the
Mn $L_{2}$ edge compared to the $L_{3}$ edge, we collected the momentum-dependent
RIXS spectra across the OO scattering peak at the Mn $L_{2}$ edge.
Figures \ref{fig:q-maps}a and \ref{fig:q-maps}b, show the momentum-dependent
RIXS maps probed by the linear $\pi$ and $\sigma$ incident polarizations,
respectively. However, the OO elastic scattering
signal completely dominates the spectra with no appreciable spectral
weight from the low energy excitations and no dependence
with the incident polarization is observed. This absence of polarization
dependence of the OO Bragg peak is consistent with the previous REXS
studies of OO in powder SBMO\citet{M.Garcia-Fernandez2008,M.Garcia-Fernandez2009,M.Garcia-Fernandez2010}. 

\begin{figure}
\includegraphics[scale=0.2]{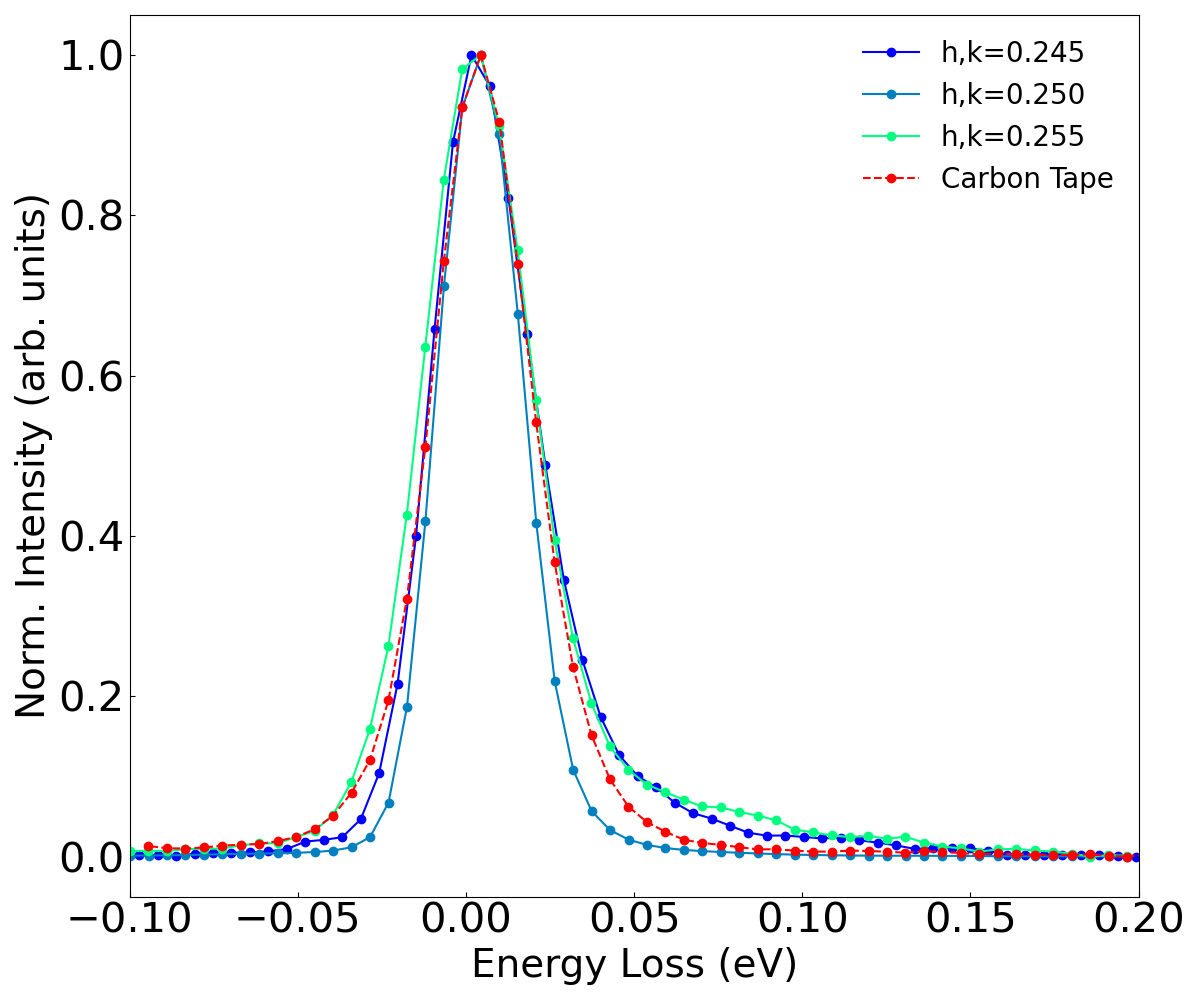}\caption{\label{fig:Line-cuts}Comparison of normalized RIXS spectra collected
at different $\mathbf{q}$ values corresponding with the grey dashed
lines in \ref{fig:q-maps}a. The data is shown together with the normalized
spectrum of a carbon tape measured at the same experimental conditions.}
\end{figure}

Figure \ref{fig:Line-cuts} shows the normalized low energy
low energy excitation spectra between 0 and 200 meV at three different
$\mathbf{q}$ positions (0.25, 0.25, 0), (0.245, 0.245, 0) and (0.255,
0.255, 0) collected at T=15K and with linear horizontal incident polarization
These $\mathbf{q}$ positions correspond to the grey dashed lines
in \ref{fig:q-maps}a and are shown together with the normalized elastic
scattering spectrum from a carbon reference sample measured at the
same geometry to illustrate the achieved energy resolution. The imperfect
tail off the carbon tape gaussian distribution reflects that the instrumental
resolution function is distorted by the optics aberration. The near
perfect gaussian shape of the OO condition is due to the strong Bragg
diffraction which suppresses the aberration. When $\mathbf{q}$ is
not exactly at the OO Bragg condition, weak low energy excitations
can be observed at around 80 meV. These data suggest that any possible
low-energy excitations are shadowed by the OO scattering at this wave
vector.

\begin{figure}
\includegraphics[scale=0.19]{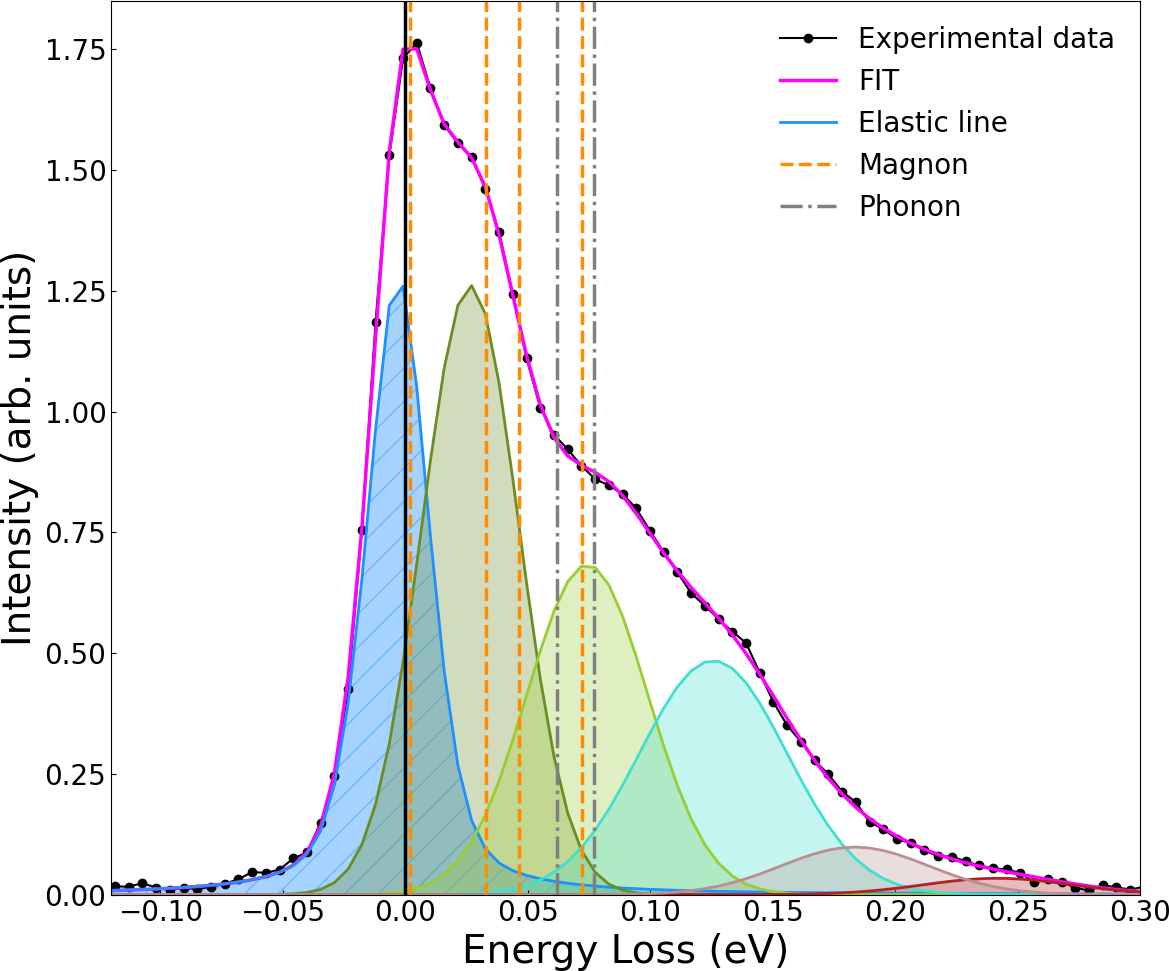}

\caption{\label{fig:excitations-with-fits}Low energy-transfer RIXS spectrum
measured at the Mn $L_{2}$ edge with $\sigma$ incident polarization
at T=15K. The $\mathbf{q}$ vector corresponds to (0.2347, 0.2347,
-0.0130) and the fits represent an elastic line and
five low energy excitations. The colors of the peaks are random as
a way to show better the individual contributions.}
\end{figure}

The RIXS data at $\mathbf{q}_{Off}$ are shown in figure \ref{fig:excitations-with-fits},
where several low-energy excitations are visible. We describe the
spectra using five gaussian functions while the elastic peak is described
with a Pseudo-Voigt function. The available energy resolution does
not allow for an univocal fit of the data and different solutions
with similar fitting errors can be achieved. However, all the tested
fit solutions point into the direction that the observed peaks are
not higher harmonics from one another. 

These results are interesting when compared to results from inelastic
neutron scattering, \citep{PhysRevB.94.014405} which studied the
magnetic excitations of the canonical half-doped manganite Pr$_{0.5}$Ca$_{0.5}$MnO$_{3}$
in its magnetic and orbitally ordered phase. The neutron study demonstrated
the existence of four gapped dispersive magnon modes. By using an
effective Heisenberg-Hamiltonian model $\mathscr{H}=-\sum_{(ij)}J_{ij}S_{i}S_{j}$,
the magnon dispersion and the dynamical structure factor were calculated.
The gapped spin wave dispersion is consistent with the Goodenough
model with strong nearest-neighbor ferromagnetic interactions along
the zigzag chains and weak antiferromagnetic interactions between
them. 

Given the same magnetic structure between PCMO and SBMO, we used the
Heisenberg Hamiltonian and the optimized exchange parameters obtained
from PCMO to mark the magnetic excitations of SBMO at the given momentum
transfer. The obtained energies for the magnon excitations for $\mathbf{q}_{Off}$
are E$_{m_{1}}=2$ meV, E$_{m_{2}}$33.09 meV, E$_{m_{3}}=46.62$
meV and E$_{m_{4}}=71.98$ meV as displayed in figure \ref{fig:excitations-with-fits}
by dashed orange lines. 

In addition to the magnons, we expect also phonons to be present in
this range of energy. Akahoshi et al. \citep{PhysRevB.70.064418}
studied SBMO using optical conductivity and Raman scattering. These
techniques are restricted to $\mathbf{q}$ = 0 and are not sensitive
to the dispersion of the observed excitations. Two major peaks are
observed in the Raman spectra around 500 and 620 cm$^{-1}$ (62 and
77 meV) that are assigned to the Jahn-Teller (JT) and breathing modes
respectively. These phonon energies are displayed in Figure 6 as dashed
grey lines.

\begin{figure}
\includegraphics[scale=0.21]{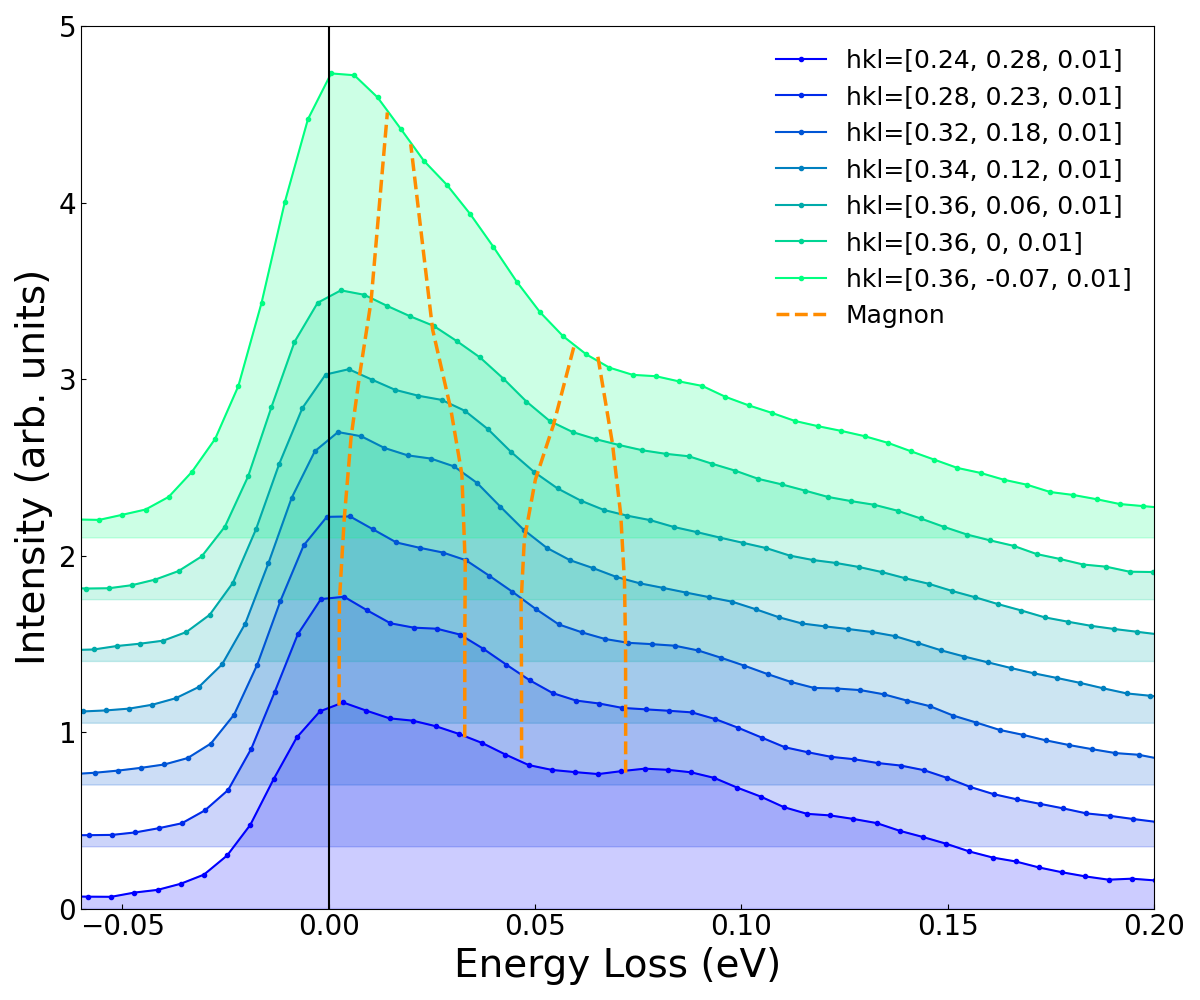}

\caption{\label{fig:q-dependence}RIXS spectra taken at the Mn $L_{2}$ edge
with $\sigma$ incident polarization at T=15K and at different $\mathbf{q}$
vectors. The dashed lines represent the dispersion of the magnon energies
for the same $\mathbf{q}$ vectors obtained from theoretical calculations.}
\end{figure}

To better understand the origin of the low energy excitations, we
probed its dispersion by collecting its momentum-dependence between
$\mathbf{Q}$ = 0.4864 and $\mathbf{Q}$ = 0.8453 with $\sigma$ incident
polarization at T=15K. The collected RIXS spectra are presented in
Figure \ref{fig:q-dependence} and are vertically displaced for clarity.
The intensity of the RIXS spectra are normalized by the d-d excitations
as these are not expected to be dispersive or have significant structure
factor modulations. The momentum-dependent low-energy modes do not
show sizable dispersion. Four dashed lines represent the calculated
dispersion of the magnetic excitations. The comparison demonstrates
that the magnons very likely contribute to the low-energy excitations
in RIXS. However, other modes, including phonons and orbitons, likely
also contribute to the RIXS spectra as the calculated magnon peaks
cannot reconcile the observed evolution of low-energy excitations.

\begin{figure}
\subfloat[]{\includegraphics[scale=0.22]{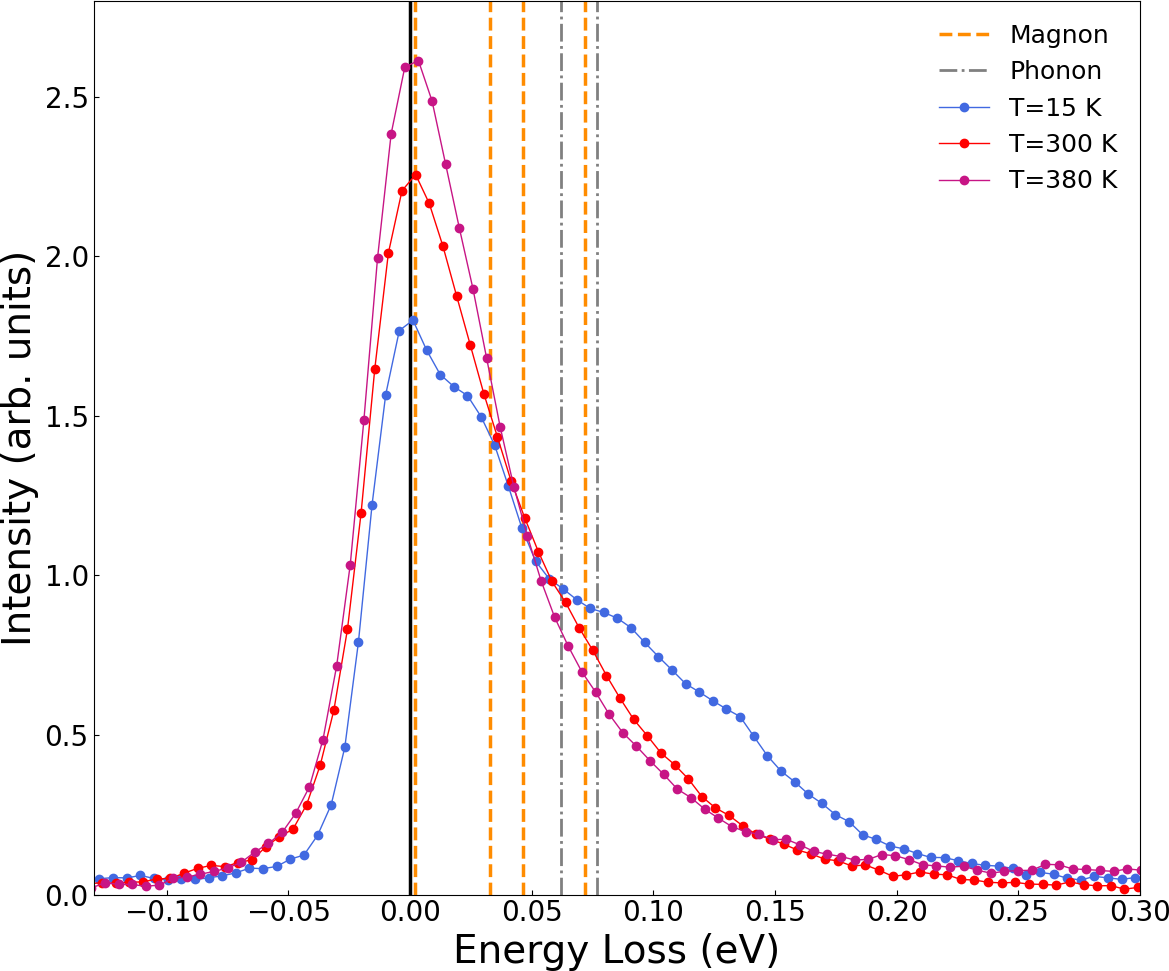}

}

\subfloat[]{\includegraphics[scale=0.21]{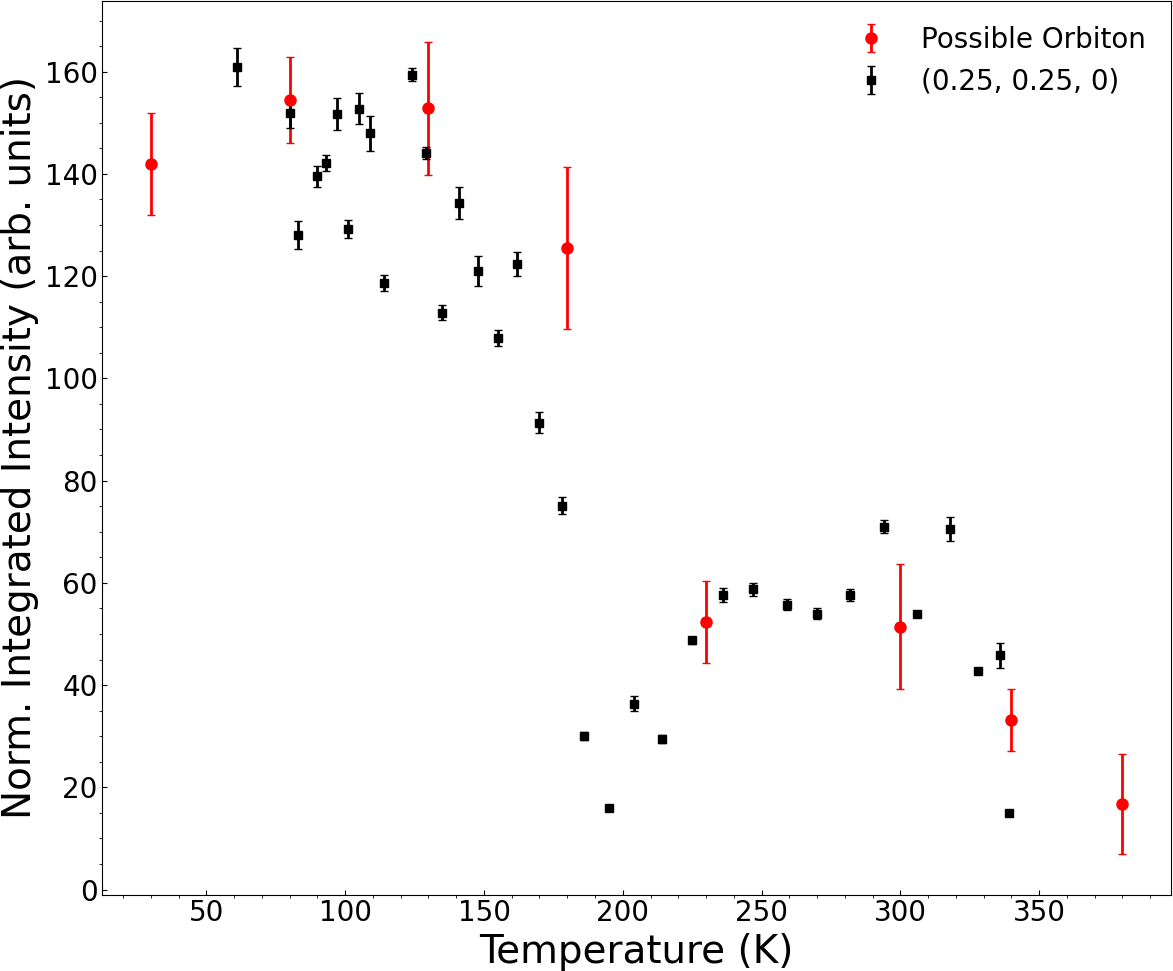}

}\caption{\label{fig:Temperature-dependence}(a) RIXS spectra measured at the
Mn $L_{2}$ edge and $\mathbf{q}_{Off}$ = (0.2347, 0.2347, -0.0130).
The spectra were collected with $\sigma$ incident polarization and
temperatures of 15 K, 300 K and 380 K. (b) Comparison of the integrated
intensity of the (0.25, 0.25, 0) orbital order peak measured by REXS
with the integrated intensity of the energy excitations between 80
and 200 meV in the RIXS spectra. }
\end{figure}

Another way to differenciate between these modes is to study their
temperature evolution. Figure \ref{fig:Temperature-dependence}a shows
the RIXS spectra at the Mn $L_{2}$ edge at $\mathbf{q}_{Off}$ for
three different temperatures. Comparing with the temperature dependence
of the magnetization recorded during cooling and heating of SBMO presented
in figure \ref{fig:Introduction}c, allows us to establish the correlation
between the RIXS signal and the phase diagram. The three chosen temperatures
are T = 15 K corresponding to the T$_{CO2}$ phase, T=300 K with T$_{CO2}$
< T < T$_{CO}$ and finally at T=380 K that is above the metal to
insulator transition T > T$_{CO}$. 

There are two regions in the spectra that exhibit opposite temperature
dependence. Below 50 meV, the intensity of the elastic and lowest
energy excitations increases with increasing temperatures. However,
for intensities above 50 meV we observe a drastic intensity reduction
up to T = 300 K and T = 380 K. The temperature evolution of the RIXS
spectra is in agreement with Raman which reported that both, the JT
and the breathing modes are rapidly suppressed with increasing temperature
towards T$_{CO2}$ and bear little intensity in the temperature range
of T$_{CO2}$ < T < T$_{CO}$. Remarkably, we observe the presence
of several excitations around 100 meV whose energies do not correspond
either to magnons or phonons. These excitations exhibit a more pronounced
temperature dependence. Ishihara et al \citep{Ishihara:2005aa}, postulated
in their theoretical study of orbital excitations in LaMnO$_{3}$,
that the orbiton energy is of the order of \textit{J} and larger than
the Jahn-Teller phonon energy. Hence, the observed excitations around
100 meV might correspond to orbital excitations. In order to test
this scenario, we collected RIXS spectra at further temperatures and
we compared its spectral weight around 100 meV with the temperature
dependence of the orbital order reflection. The RIXS spectra was fitted
using a skewed Gaussian peak to describe the elastic signal and two
Gaussian peaks to describe the inelastic low energy excitations. The
first Gaussian peak was used to describe the excitations below 50
meV, that as \ref{fig:Temperature-dependence}a shows increase in
intensity with increasing temperature. The second Gaussian peak describes
the energy region between 50 meV and 200 meV that exhibits a decreasing
spectral weight as function of raising temperature. Figure \ref{fig:Temperature-dependence}b
compares the integrated intensity of the REXS signal
as function of temperature with the integrated intensity of the inelastic
excitations between 50 meV and 200 meV. A remarkable agreement in
their temperature dependence can be observed.

One of the specific properties of the A-site ordered manganites is
the transition at T$_{CO2}$ = 210 K associated with a change in the
stacking of the orbitally ordered planes along the z axis. In the
OO temperature dependence it results in a steep increase of intensity
by a factor of two. This peculiar behavior was already observed in
a previous REXS study of SBMO powder samples \citep{M.Garcia-Fernandez2008}
and appears also in our REXS measurements of this single crystal.
Remarkably, the temperature dependence of the integrated intensity
of the excitations around 100 meV follows also this behavior, including
the doubling of intensity compared to below T$_{CO2}$ which indicates
a correlation between orbital order and the excitations. This evolution
is in strong contrast with the one of the excitations at energies
below 50 meV.

Several works have considered orbital excitations in the parent compound
LaMnO$_{3}$\citep{Brink:2001aa,Allen:1999aa,Schmidt:2007aa,PhysRevLett.101.106406},
with significant variation of the role attributed to phonons. Van
der Brink\citep{Brink:2001aa} used self-consistent second order perturbation
theory in the orbiton-phonon coupling and found that the orbiton dispersion
is strongly reduced by the electron-phonon coupling. In addition,
he found that this coupling also mixes the orbiton and phonon modes
and causes satellite structures in the orbiton and the phonon spectral
function. In this context the observed excitations
between 80 and 200 meV could be explained as the orbital collective
excitations plus these phonon satellites whose energy would involve
contributions from both the orbital and the lattice components.

Schmidt et al. \citep{Schmidt:2007aa} performed a calculation to
provide information on the nature of orbital excitations in the presence
of substantial orbiton phonon coupling. In their model, they considered
quantitatively the dynamic orbiton-phonon interaction that corresponds
to the creation or annihilation of a local distortive phonon in the
presence of an orbiton. Their results show that the system becomes
more local when increasing the orbiton-phonon coupling, i.e., the
effective bandwidth of the orbiton band decreases as the orbiton-phonon
coupling increases. In a later work, that focussed completely on the
observation of orbital excitations in LaMnO$_{3}$ using RIXS, Forte
et al.\citep{PhysRevLett.101.106406} used the ultra short core-hole
lifetime expansion for RIXS and obtained a gapped orbiton spectrum.
Since their calculated orbital Hamiltonian did not have a continuous
symmetry, the Goldstone models were absent. In addition they found
that at high symmetry points in the Brillouin zone, the intensity
of specific orbiton branches vanishes. This result is consistent with
our experimental data that shows that at the OO $\mathbf{q}$ vector
the OO diffraction peak dominates the spectra. 

We are aware that these experimental observations are not sufficient
to univocally claim that the observed excitations correspond to the
elusive orbitons. However, the observed correlation is a strong motivation
for further studies. We believe that in order to properly explore
the low energy excitations in half doped manganites, further improved
energy resolution and additional polarization analysis of the scattered
radiation would be essential to clarify the electronic nature of the
observed excitations. As it would allow us to distinguish
the excitations that are present in each of the four polarization
channels. Note that for coupled orbital-phonon modes, one can only
see the 'orbital' contributions as is the case for the measurement
of phonons at the oxygen K-edge, see e.g. Ueda et al.\citet{Ueda:2023aa}.

\section{Conclusions}

Thanks to the improved flux and energy resolution of new generation
soft x-ray RIXS instruments, a re-examination of low energy excitations
in half doped A-site ordered SBMO combining RIXS and REXS measurements
was performed. At the OO wave vector the OO diffraction peak dominates
the spectra completely. When moving slightly away from the Bragg condition,
several low energy excitations became observable.

The observed low energy excitations fall within the energy region
in which both magnetic and phonon excitations are expected to appear.
Comparing to neutron inelastic and Raman scattering results we identified
the energies at which magnons and phonons would be expected. We find
additional excitations between 80 and 200 meV that could not be identified
as magnons or phonons. They could potentially correspond to the elusive
orbiton. The temperature dependence of the RIXS and REXS signals shows
a clear correlation for the energy excitations between 80 and 200
meV. The unique temperature dependence indicates a clear correlation
between orbital order and the excitations. This observed correlation
hints to the possibility that these excitations could correspond to
the long sought orbiton. 
\begin{acknowledgments}
We would like to thank Dr. Russel Ewings for the discussions about
the neutron data. We thank Diamond Light Source for the provision
of beamtime on beamline I21 under proposals mm24600-1 and mm30866-1
and for access to the facilities of the Materials Characterization
Laboratory.
\end{acknowledgments}

\bibliography{SBMOcopy}

@article{Ueda:2023aa,
	author = {Ueda, Hiroki and Garc{\'\i}a-Fern{\'a}ndez, Mirian and Agrestini, Stefano and Romao, Carl P. and van den Brink, Jeroen and Spaldin, Nicola A. and Zhou, Ke-Jin and Staub, Urs},
	journal = {Nature},
	number = {7967},
	pages = {946--950},
	title = {Chiral phonons in quartz probed by X-rays},
	volume = {618},
	year = {2023}}

@article{Staub:2009aa,
	author = {Staub, U. and Garc{\'\i}a-Fern{\'a}ndez, M. and Bodenthin, Y. and Scagnoli, V. and De Souza, R. A. and Garganourakis, M. and Pomjakushina, E. and Conder, K.},
	journal = {Physical Review B},
	month = {06},
	number = {22},
	pages = {224419--},
	title = {Orbital and magnetic ordering in ${\text{Pr}}_{1\ensuremath{-}x}{\text{Ca}}_{x}{\text{MnO}}_{3}$ and ${\text{Nd}}_{1\ensuremath{-}x}{\text{Sr}}_{x}{\text{MnO}}_{3}$ manganites near half doping studied by resonant soft x-ray powder diffraction},
	volume = {79},
	year = {2009}}

@article{Beale:2009aa,
	author = {Beale, T. A. W. and Bland, S. R. and Johnson, R. D. and Hatton, P. D. and Cezar, J. C. and Dhesi, S. S. and v. Zimmermann, M. and Prabhakaran, D. and Boothroyd, A. T.},
	journal = {Physical Review B},
	month = {02},
	number = {5},
	pages = {054433--},
	title = {Thermally induced rotation of 3d orbital stripes in PrSr0.1Ca0.9MnO7},
	volume = {79},
	year = {2009}}

@article{Wilkins:2003aa,
	author = {Wilkins, S. B. and Spencer, P. D. and Hatton, P. D. and Collins, S. P. and Roper, M. D. and Prabhakaran, D. and Boothroyd, A. T.},
	journal = {Physical Review Letters},
	month = {10},
	number = {16},
	pages = {167205--},
	title = {Direct Observation of Orbital Ordering in La0.5Sr1.5MnO4 Using Soft X-ray Diffraction},
	volume = {91},
	year = {2003}}

@article{Dhesi:2004aa,
	author = {Dhesi, S. S. and Mirone, A. and De Nada{\"\i}, C. and Ohresser, P. and Bencok, P. and Brookes, N. B. and Reutler, P. and Revcolevschi, A. and Tagliaferri, A. and Toulemonde, O. and van der Laan, G.},
	journal = {Physical Review Letters},
	month = {02},
	number = {5},
	pages = {056403--},
	title = {Unraveling Orbital Ordering in La0.5Sr1.5MnO4},
	volume = {92},
	year = {2004}}

@article{Murakami:1998aa,
	author = {Murakami, Y. and Kawada, H. and Kawata, H. and Tanaka, M. and Arima, T. and Moritomo, Y. and Tokura, Y.},
	journal = {Physical Review Letters},
	month = {03},
	number = {9},
	pages = {1932--1935},
	title = {Direct Observation of Charge and Orbital Ordering in La0.5Sr1.5MnO4},
	volume = {80},
	year = {1998}}

@article{Staub:2005aa,
	author = {Staub, U. and Scagnoli, V. and Mulders, A. M. and Katsumata, K. and Honda, Z. and Grimmer, H. and Horisberger, M. and Tonnerre, J. M.},
	journal = {Physical Review B},
	month = {06},
	number = {21},
	pages = {214421--},
	title = {Orbital and magnetic ordering in La0.5Sr1.5MnO4 studied by soft x-ray resonant scattering},
	volume = {71},
	year = {2005}}

@article{Thomas:2004aa,
	author = {Thomas, K. J. and Hill, J. P. and Grenier, S. and Kim, Y-J. and Abbamonte, P. and Venema, L. and Rusydi, A. and Tomioka, Y. and Tokura, Y. and McMorrow, D. F. and Sawatzky, G. and van Veenendaal, M.},
	journal = {Physical Review Letters},
	month = {06},
	number = {23},
	pages = {237204--},
	title = {Soft X-Ray Resonant Diffraction Study of Magnetic and Orbital Correlations in a Manganite Near Half Doping},
	volume = {92},
	year = {2004}}

@article{Allen:1999aa,
	author = {Allen, Philip B. and Perebeinos, Vasili},
	journal = {Physical Review Letters},
	month = {12},
	number = {23},
	pages = {4828--4831},
	title = {Self-Trapped Exciton and Franck-Condon Spectra Predicted in LaMnO3},
	volume = {83},
	year = {1999}}

@article{Schmidt:2007aa,
	author = {Schmidt, K. P. and Gr{\"u}ninger, M. and Uhrig, G. S.},
	journal = {Physical Review B},
	month = {08},
	number = {7},
	pages = {075108--},
	title = {Fate of orbitons coupled to phonons},
	volume = {76},
	year = {2007}}

@article{Polli:2007aa,
	author = {Polli, D. and Rini, M. and Wall, S. and Schoenlein, R. W. and Tomioka, Y. and Tokura, Y. and Cerullo, G. and Cavalleri, A.},
	journal = {Nature Materials},
	number = {9},
	pages = {643--647},
	title = {Coherent orbital waves in the photo-induced insulator--metal dynamics of a magnetoresistive manganite},
	volume = {6},
	year = {2007}}

@article{Gruninger:2002aa,
	author = {Gr{\"u}ninger, M. and R{\"u}ckamp, R. and Windt, M. and Reutler, P. and Zobel, C. and Lorenz, T. and Freimuth, A. and Revcolevschi, A.},
	journal = {Nature},
	number = {6893},
	pages = {39--40},
	title = {Experimental quest for orbital waves},
	volume = {418},
	year = {2002}}

@article{Saitoh:2001aa,
	author = {Saitoh, E. and Okamoto, S. and Takahashi, K. T. and Tobe, K. and Yamamoto, K. and Kimura, T. and Ishihara, S. and Maekawa, S. and Tokura, Y.},
	journal = {Nature},
	number = {6825},
	pages = {180--183},
	title = {Observation of orbital waves as elementary excitations in a solid},
	volume = {410},
	year = {2001}}

@article{Brink:1999aa,
	author = {van den Brink, Jeroen and Khaliullin, Giniyat and Khomskii, Daniel},
	journal = {Physical Review Letters},
	month = {12},
	number = {24},
	pages = {5118--5121},
	title = {Charge and Orbital Order in Half-Doped Manganites},
	volume = {83},
	year = {1999}}

@article{Ishihara:2005aa,
	author = {Ishihara, Sumio and Murakami, Youichi and Inami, Toshiya and Ishii, Kenji and Mizuki, Jun'ichiro and Hirota, Kazuma and Maekawa, Sadamichi and Endoh, Yasuo},
	journal = {New Journal of Physics},
	number = {1},
	pages = {119},
	title = {Theory and experiment of orbital excitations in correlated oxides},
	volume = {7},
	year = {2005}}

@article{Brink:2001aa,
	author = {van den Brink, Jeroen},
	journal = {Physical Review Letters},
	month = {11},
	number = {21},
	pages = {217202--},
	title = {Orbital Excitations in LaMnO3},
	volume = {87},
	year = {2001}}

@article{Schlappa:2012aa,
	author = {Schlappa, J. and Wohlfeld, K. and Zhou, K. J. and Mourigal, M. and Haverkort, M. W. and Strocov, V. N. and Hozoi, L. and Monney, C. and Nishimoto, S. and Singh, S. and Revcolevschi, A. and Caux, J. -S. and Patthey, L. and R{\o}nnow, H. M. and van den Brink, J. and Schmitt, T.},
	journal = {Nature},
	number = {7396},
	pages = {82--85},
	title = {Spin--orbital separation in the quasi-one-dimensional Mott insulator Sr2CuO3},
	volume = {485},
	year = {2012}}

@article{PhysRevB.94.014405,
	author = {Ewings, R. A. and Perring, T. G. and Sikora, O. and Abernathy, D. L. and Tomioka, Y. and Tokura, Y.},
	journal = {Phys. Rev. B},
	month = {Jul},
	pages = {014405},
	title = {Spin excitations used to probe the nature of exchange coupling in the magnetically ordered ground state of Pr0.5Ca0.5MnO3},
	volume = {94},
	year = {2016}}

@article{Zhou:rv5159,
	author = {Zhou, Ke-Jin and Walters, Andrew and Garcia-Fernandez, Mirian and Rice, Thomas and Hand, Matthew and Nag, Abhishek and Li, Jiemin and Agrestini, Stefano and Garland, Peter and Wang, Hongchang and Alcock, Simon and Nistea, Ioana and Nutter, Brian and Rubies, Nicholas and Knap, Giles and Gaughran, Martin and Yuan, Fajin and Chang, Peter and Emmins, John and Howell, George},
	journal = {Journal of Synchrotron Radiation},
	month = {Mar},
	number = {2},
	pages = {563--580},
	title = {I21: an advanced high-resolution resonant inelastic X-ray scattering beamline at Diamond Light Source},
	volume = {29},
	year = {2022}}

@article{PhysRevLett.94.047203,
	author = {Grenier, S. and Hill, J. P. and Kiryukhin, V. and Ku, W. and Kim, Y.-J. and Thomas, K. J. and Cheong, S-W. and Tokura, Y. and Tomioka, Y. and Casa, D. and Gog, T.},
	journal = {Phys. Rev. Lett.},
	month = {Feb},
	pages = {047203},
	title = {$d\mathrm{\text{\ensuremath{-}}}d$ Excitations in Manganites Probed by Resonant Inelastic X-Ray Scattering},
	volume = {94},
	year = {2005}}

@article{PhysRevB.70.224437,
	author = {Ishii, K. and Inami, T. and Ohwada, K. and Kuzushita, K. and Mizuki, J. and Murakami, Y. and Ishihara, S. and Endoh, Y. and Maekawa, S. and Hirota, K. and Moritomo, Y.},
	journal = {Phys. Rev. B},
	month = {Dec},
	pages = {224437},
	title = {Resonant inelastic x-ray scattering study of the hole-doped manganites ${\mathrm{La}}_{1\ensuremath{-}x}{\mathrm{Sr}}_{x}{\mathrm{MnO}}_{3}$ ($x=0.2$, 0.4)},
	volume = {70},
	year = {2004}}

@article{PhysRevB.67.045108,
	author = {Inami, T. and Fukuda, T. and Mizuki, J. and Ishihara, S. and Kondo, H. and Nakao, H. and Matsumura, T. and Hirota, K. and Murakami, Y. and Maekawa, S. and Endoh, Y.},
	journal = {Phys. Rev. B},
	month = {Jan},
	pages = {045108},
	title = {Orbital excitations in ${\mathrm{LaMnO}}_{3}$ studied by resonant inelastic x-ray scattering},
	volume = {67},
	year = {2003}}

@article{PhysRevB.64.014414,
	author = {Kondo, H. and Ishihara, S. and Maekawa, S.},
	journal = {Phys. Rev. B},
	month = {Jun},
	pages = {014414},
	title = {Resonant inelastic x-ray scattering from charge and orbital excitations in manganites},
	volume = {64},
	year = {2001}}

@article{PhysRevLett.101.106406,
	author = {Forte, Filomena and Ament, Luuk J. P. and van den Brink, Jeroen},
	journal = {Phys. Rev. Lett.},
	month = {Sep},
	pages = {106406},
	title = {Single and Double Orbital Excitations Probed by Resonant Inelastic X-Ray Scattering},
	volume = {101},
	year = {2008}}

@article{ISHIHARA200415,
	author = {S. Ishihara and H. Kondoh and S. Maekawa},
	journal = {Physica B: Condensed Matter},
	number = {1},
	pages = {15-18},
	title = {Resonant inelastic X-ray scattering in manganites with perovskite structure},
	volume = {345},
	year = {2004}}

@article{PhysRevB.62.2338,
	author = {Ishihara, Sumio and Maekawa, Sadamichi},
	journal = {Phys. Rev. B},
	month = {Jul},
	pages = {2338--2345},
	title = {Theory of orbital excitation and resonant inelastic x-ray scattering in manganites},
	volume = {62},
	year = {2000}}

@article{Tokunaga:2006aa,
	author = {Tokunaga, Yusuke and Lottermoser, Thomas and Lee, Yunsang and Kumai, Reiji and Uchida, Masaya and Arima, Takahisa and Tokura, Yoshinori},
	journal = {Nature Materials},
	number = {12},
	pages = {937--941},
	title = {Rotation of orbital stripes and the consequent charge-polarized state in bilayer manganites},
	volume = {5},
	year = {2006}}

@article{doi:10.1143/JPSJ.81.093602,
	author = {Morikawa ,Daisuke and Tsuda ,Kenji and Maeda ,Youichi and Yamada ,Shigeki and Arima ,Taka-hisa},
	journal = {Journal of the Physical Society of Japan},
	number = {9},
	pages = {093602},
	title = {Charge and Orbital Order Patterns in an A-Site Ordered Perovskite-Type Manganite SmBaMn2O6 Determined by Convergent-Beam Electron Diffraction},
	volume = {81},
	year = {2012}}

@article{PhysRevB.70.064418,
	author = {Akahoshi, D. and Okimoto, Y. and Kubota, M. and Kumai, R. and Arima, T. and Tomioka, Y. and Tokura, Y.},
	journal = {Phys. Rev. B},
	month = {Aug},
	pages = {064418},
	title = {Charge-orbital ordering near the multicritical point in $A$-site ordered perovskites $\mathrm{Sm}\mathrm{Ba}{\mathrm{Mn}}_{2}{\mathrm{O}}_{6}$ and $\mathrm{Nd}\mathrm{Ba}{\mathrm{Mn}}_{2}{\mathrm{O}}_{6}$},
	volume = {70},
	year = {2004}}

@article{PhysRevB.66.140408,
	author = {Arima, T. and Akahoshi, D. and Oikawa, K. and Kamiyama, T. and Uchida, M. and Matsui, Y. and Tokura, Y.},
	journal = {Phys. Rev. B},
	month = {Oct},
	pages = {140408},
	title = {Change in charge and orbital alignment upon antiferromagnetic transition in the A-site-ordered perovskite manganese oxide $R{\mathrm{BaMn}}_{2}{\mathrm{O}}_{6}$ $(R=\mathrm{Tb}$ and Sm)},
	volume = {66},
	year = {2002}}

@article{NAKAJIMA2004987,
	author = {T Nakajima and H Kageyama and M Ichihara and K Ohoyama and H Yoshizawa and Y Ueda},
	journal = {Journal of Solid State Chemistry},
	number = {3},
	pages = {987-999},
	title = {Anomalous octahedral distortion and multiple phase transitions in the metal-ordered manganite YBaMn2O6},
	volume = {177},
	year = {2004}}

@article{doi:10.1143/JPSJ.72.241,
	author = {Kageyama ,H. and Nakajima ,T. and Ichihara ,M. and Ueda ,Y. and Yoshizawa ,H. and Ohoyama ,K.},
	journal = {Journal of the Physical Society of Japan},
	number = {2},
	pages = {241-244},
	title = {New Stacking Variations of the Charge and Orbital Ordering in the Metal-Ordered Manganite YBaMn2O6},
	volume = {72},
	year = {2003}}

@article{doi:10.1143/JPSJ.71.2605,
	author = {Uchida ,Masaya and Akahoshi ,Daisuke and Kumai ,Reiji and Tomioka ,Yasuhide and Arima ,Taka-hisa and Tokura ,Yoshinori and Matsui ,Yoshio},
	journal = {Journal of the Physical Society of Japan},
	number = {11},
	pages = {2605-2608},
	title = {Charge/Orbital Ordering Structure in Ordered Perovskite Sm1/2Ba1/2MnO3},
	volume = {71},
	year = {2002}}

@article{doi:10.1143/JPSJ.73.2283,
	author = {Nakajima ,Tomohiko and Yoshizawa ,Hideki and Ueda ,Yutaka},
	journal = {Journal of the Physical Society of Japan},
	number = {8},
	pages = {2283-2291},
	title = {A-site Randomness Effect on Structural and Physical Properties of Ba-based Perovskite Manganites},
	volume = {73},
	year = {2004}}

@article{M.Garcia-Fernandez2009,
	author = {Garc\'{\i}a-Fern\'andez, M. and Staub, U. and Bodenthin, Y. and Scagnoli, V. and Pomjakushin, V. and Lovesey, S. W. and Mirone, A. and Herrero-Mart\'{\i}n, J. and Piamonteze, C. and Pomjakushina, E.},
	journal = {Phys. Rev. Lett.},
	month = {Aug},
	pages = {097205},
	title = {Orbital Order at Mn and O Sites and Absence of Zener Polaron Formation in Manganites},
	volume = {103},
	year = {2009}}

@article{M.Garcia-Fernandez2010,
	author = {Garc\'{\i}a-Fern\'andez, M. and Staub, U. and Bodenthin, Y. and Pomjakushin, V. and Mirone, A. and Fern\'andez-Rodr\'{\i}guez, J. and Scagnoli, V. and Mulders, A. M. and Lawrence, S. M. and Pomjakushina, E.},
	journal = {Phys. Rev. B},
	month = {Dec},
	pages = {235108},
	title = {Doping and temperature dependence of $\text{Mn}\text{ }3d$ states in $A$-site ordered manganites},
	volume = {82},
	year = {2010}}

@article{M.Garcia-Fernandez2008,
	author = {Garc\'{\i}a-Fern\'andez, M. and Staub, U. and Bodenthin, Y. and Lawrence, S. M. and Mulders, A. M. and Buckley, C. E. and Weyeneth, S. and Pomjakushina, E. and Conder, K.},
	journal = {Phys. Rev. B},
	month = {Feb},
	pages = {060402},
	title = {Resonant soft x-ray powder diffraction study to determine the orbital ordering in A-site-ordered ${\mathrm{SmBaMn}}_{2}{\mathrm{O}}_{6}$},
	volume = {77},
	year = {2008}}

\end{document}